\documentclass[twocolumn,secnumarabic,amssymb, amsmath, nofootinbib,tightenlines,
nobibnotes, aps, prl,epsfig]{revtex4}
\usepackage{graphicx}
\usepackage{dcolumn}
\usepackage{bm}
\begin{document}
\preprint{APS/123-QED}
\title{Color dipole cross section in the DGLAP improved saturation model}

\author{G.R.Boroun}%
 \email{grboroun@gmail.com; boroun@razi.ac.ir }
\affiliation{ Physics Department, Razi University, Kermanshah
67149, Iran}
\date{\today}
\begin{abstract}
We show that the geometric scaling of the  dipole cross section can be explained using standard
 DGLAP perturbative evolution.
 The DGLAP improved saturation model due to the  Laplace transforms
 method is considered at LO and NNLO approximations from the experimental data by
relying on a Froissart-bounded parametrization of
$F_{2}(x,Q^{2})$. These results
 are comparable with Golec-Biernat-W$\ddot{\mathrm{u}}$sthoff (GBW)
 model in a wide kinematic region $rQ_{s}$ which takes into account charm mass. The successful
 description of ${\sigma}_{\mathrm{dip}}(x,r)/\sigma_{0}$ and ${\sigma}_{\mathrm{dip}}(rQ_{s})/\sigma_{0}$ are presented.\\

\end{abstract}
 \pacs{***}
\keywords{****} 
\maketitle
\subsection{1. Introduction}

Recently an update [1] on the saturation model of deep inelastic
scattering (DIS) by Golec-Biernat $et~al$. is presented by
introducing the results of new fits [2] to the extracted HERA data
[3] on the proton structure function at small $x$ with the GBW
saturation model and its modification to cover high values of
$Q^{2}$. When $x{\ll}1$, the DGLAP [4] or the BFKL [5] evolution
equations predict that the small $x$ structure of the proton is
dominated by a strongly rising gluon density, which drives a
similar rise of the sea quark densities. In this region gluons in
the proton form a dense system with mutual interaction and
recombination which leads to the saturation of the total cross
section [5]. For $x{\approx}Q^{2}/W^{2}{\ll}1$, the virtual
spacelike photon on the proton fluctuates are defined into
on-shell quark-antiquark, $q\overline{q}$, vector state. Here
$Q^{2}$ refers to the photon virtuality and $W$ to the
photon-proton center-of-mass energy. In this process  photon
interact with the proton via coupling of two gluons to the
$q\overline{q}$ color dipole, where this called the color dipole
model (CDM). The mass of $q\overline{q}$ dipole, in terms of the
transverse momentum $\overrightarrow{k_{\bot}}$ is given by
$M^{2}_{q\overline{q}}=\frac{\overrightarrow{k_{\bot}}^{2}}{z(1-z)}$,
where $\overrightarrow{k_{\bot}}$ is defined with respect to the
photon direction and the variable $z$ characterizes the
distribution of the momenta between quark and antiquark [6]. The
lifetime of the $q\overline{q}$ dipole is defined by
$\tau=\frac{W^{2}}{Q^{2}+M^{2}_{q\overline{q}}}\gg
\frac{1}{M_{p}}$, which it is much longer than its typical
interaction time with the target at  small $x$. This condition not
only restricts the kinematical range of the color dipole model to
$x\ll 1$ but also saturate the $\gamma^{*}$-proton cross section
with $x<0.1$ [7].\\
Some years ago [8] the saturation model was shown by Golec-Biernat
and  W$\ddot{\mathrm{u}}$sthoff which give an elegant and accurate
account of DIS at small $x$ and has been formulated to new models
 in recent years [9-13]. This type saturation occurs when the
 photon wavelength $1/Q$ reaches the size of the proton. It is
 well known that the dipole picture is a factorization scheme for
 DIS, which is particularly convenient for the inclusion of
 unitarity corrections at small $x$. In the mixed representation, the scattering between
 the virtual photon $\gamma^{*}$ and the proton is seen as the
 color dipole where the transverse dipole size $r$ and the
 longitudinal momentum fraction $z$ with respect to the photon
 momentum are defined. The amplitude for the complete process is simply the produce of
these subprocess amplitudes, as the DIS cross section is
factorized into a light-cone wave function and a dipole cross
section. Using the optical theorem, this leads to the following
expression for the $\gamma^{*}p$ cross-sections
\begin{eqnarray}
\sigma_{L,T}^{\gamma^{*}p}(x,Q^{2})=\int dz d^{2}\mathbf{r}
|\Psi_{L,T}(\mathbf{r},z,Q^{2})|^{2}\sigma_{\mathrm{dipole}}(\widetilde{x}_{f},\mathbf{r}),
\end{eqnarray}
and the $F_{2}$ structure function is defined as
\begin{eqnarray}
F_{2}(x,Q^{2})&=&\frac{Q^{2}}{4\pi^{2}\alpha}[\sigma_{L}^{\gamma^{*}p}(x,Q^{2})+\sigma_{T}^{\gamma^{*}p}(x,Q^{2})].
\end{eqnarray}
The subscript $L$ and  $T$ referring to the transverse and
longitudinal polarization state of the exchanged boson. Here
$\Psi_{L,T}$ are the appropriate spin averaged light-cone wave
functions of the photon and
$\sigma_{\mathrm{dip}}(\widetilde{x}_{f},r)$ is the dipole
cross-section which related to the imaginary part of the
$(q\overline{q})p$ forward scattering amplitude and
$\widetilde{x}_{f}{\equiv}x(1+4m_{f}^{2}/Q^{2})$ is equivalent to
the Bjorken variable and provides an interpolation for the
$Q^{2}{\rightarrow}0$ limit, $m_{f}$ is the mass of the quark of
flavour $f$. The variable $z$, with $0\leq z \leq 1 $,
characterizes the distribution of the momenta between quark and
antiquark. The square of the photon wave function describes the
probability for the occurrence of a $(q\overline{q})$ fluctuation
of transverse size with respect to
the photon polarization [1,8-15].\\
The dipole hadron cross section $\sigma_{\mathrm{dip}}$ contains
all information about the target and the strong interaction
physics. There are several phenomenological implementations for
this quantity and the main feature is to be able to match the soft
(low $Q^{2}$) and hard (large $Q^{2}$) regimes in an unified way.
In Ref.[8], the dipole cross section was proposed to have the
eikonal-like form
\begin{eqnarray}
\sigma_{\mathrm{dip}}(\widetilde{x}_{f},r)=\sigma_{0}(1-e^{-r^{2}Q^{2}_{s}/4}),
\end{eqnarray}
where $Q_{s}(\widetilde{x})$ plays the role of the saturation
momentum, parametrized as
$Q^{2}_{s}(x)=Q_{0}^{2}(\widetilde{x}/x_{0})^{-\lambda}$.
Parameters $Q_{0}$ and $x_{0}$ set dimension and absolute value of
the saturation scale and exponent $\lambda$ governs $x$ behavior
of $Q_{s}^{2}$. The saturation (non-linear QCD) is energy
dependent and marks the transition between the linear (leading
twist) perturbative QCD regime and saturation domain. The
resulting dipole cross section presents the colour transparency
property, i.e. $\sigma_{\mathrm{dip}}\sim r^{2}$ when
$r{\rightarrow}0$, which is purely pQCD phenomenon and the
saturation property, i.e. $\sigma_{\mathrm{dip}}\sim \sigma_{0}$
at large $r$, which imposes the unitarity condition. The GBW model
was updated in [10,16] to improve the large $Q^{2}$ description of
$F_{2}$ by a modification of the small $r$ behavior of the dipole
cross section to include the DGLAP evolved gluon distribution. A
similar in spirit parameterization of the dipole scattering
amplitude, based on the Balitsky-Kovchegov (BK) equation solution,
was proposed in [9]. The BK equation [17] for a dipole scattering
amplitude was proposed in terms of the hierarchy of equations for
Wilson line operators in the limit of large number of colors
$N_{c}$. The geometrical scaling (GS) [18] at the high-energy
limit of perturbative QCD obtained from the BK equation [17] and
the Colour Glass Condensate (CGC) formalism [19]. Geometrical
scaling is connected to the existence of the saturation scale and
is defined as dependence of the dipole cross section only on one
dimensionless variable.\\
In the limit of large $Q^{2}$ values, the structure function (2)
does not exactly match with the DGLAP formula for $F_{2}$, i.e.
the saturation model does not include logarithmic scaling
violations. Since the energy dependence in large $Q^{2}$ region is
mainly due to the behavior of the dipole cross section at small
dipole size $r$, therefore authors in Ref.[10] investigated the
DGLAP evolution for small dipoles. Bartels-Golec-Bienat-Kowalski
(BGBK) improved the dipole cross section by adding the collinear
DGLAP effects. Indeed the BGBK model is the implementation of QCD
evolution in the the dipole cross section which depends on the
gluon distribution. The following modification of the DGLAP
improved saturation model [1] proposed for the dipole cross
section as
\begin{eqnarray}
\sigma_{\mathrm{dip}}=\sigma_{0}\{1-\exp(-\frac{\pi^{2}r^{2}\alpha_{s}(\mu^{2})G(\widetilde{x},\mu^{2})}{3\sigma_{0}})\},
\end{eqnarray}
where the hard scale is  assumed to have the form
\begin{eqnarray}
\mu^{2}=C/r^{2}+\mu^{2}_{0},
\end{eqnarray}
and the parameters $C$ and $\mu^{2}_{0}$ are obtained from the fit
to the DIS data [1]. The gluon distribution $G(x,\mu^{2})$ obeys
the DGLAP evolution equation truncated to the gluonic sector, as
reported in literatures [1,8-19], by the following form
\begin{eqnarray}
\frac{{\partial}g(x,\mu^{2})}{{\partial}{\ln}\mu^{2}}=\frac{\alpha_{s}(\mu^{2})}{2\pi}{\int_{x}^{1}}\frac{dz}{z}P_{gg}(z)g(\frac{x}{z},\mu^{2}),
\end{eqnarray}
where $g(x,\mu^{2})$ is the gluon density and
$G(x,\mu^{2})=xg(x,\mu^{2})$. The splitting function $P_{gg}$ at
the leading-order (LO) approximation reads
\begin{eqnarray}
P^{\rm LO}_{gg}(z)&=&2C_{A}(\frac{z}{(1-z)_{+}}+\frac{(1-z)}{z}+z(1-z))\nonumber\\
&&+\delta(1-z)\frac{(11C_{A}-4n_{f}T_{R})}{6},
\end{eqnarray}
with $C_{A}=N_{c}=3$,
$C_{F}=\frac{N_{c}^{2}-1}{2N_{c}}=\frac{4}{3}$ and
 $T_{f}=\frac{1}{2}n_{f}$ where $n_{f}$ is the active quark flavor. The convolution integrals in (6) which
contains plus prescription, $()_{+}$, can be easily calculate by
\begin{eqnarray}
\int_{x}^{1}\frac{dy}{y}f(\frac{x}{y})_{+}g(y)&=&\int_{x}^{1}\frac{dy}{y}f(\frac{x}{y})[g(y)-\frac{x}{y}g(x)]\nonumber\\
&&-g(x)\int_{0}^{x}f(y)dy.
\end{eqnarray}
The initial gluon distribution is defined at the scale
$\mu_{0}^{2}$ in the form [1]
\begin{eqnarray}
xg(x,\mu_{0}^{2})=A_{g}x^{-\lambda_{g}}(1-x)^{5.6}.
\end{eqnarray}
The choice of the power 5.6, which regulates the large-$x$
behavior, and another parameters (i.e., $A_{g}$ and $\lambda_{g}$)
are motivated by global fits to DIS data with the LO DGLAP
equation in literatures.\\
Although BGBK model is  successful in describing dipole cross
section at large values of $r$ as the two models (GBW and BGBK)
overlap in this region but they differ in the small $r$ region
where the running of the gluon distribution starts to play a
significant role. Indeed the DGLAP improved model of
$\sigma_{\mathrm{dip}}$ significantly improves agreement at large
values of $Q^{2}$ without affecting the physics of saturation
responsible for transition to small $Q^{2}$. As expected, GS is
true for the DGLAP improved model curve for the scaling variable
$rQ_{s}{\geq}1$ and for the GBW model curve for the whole region
[1].\\
It is well known that the color dipole cross sections determined
from the original structure functions with a parametrization of
the deep inelastic structure function for electromagnetic
scattering with protons in Ref.[20]. The authors in Ref.[20]
presented the dipole cross section from an approximate form of the
presumed dipole cross section convoluted with the perturbative
photon wave function for virtual photon splitting into a color
dipole with massless quarks. Some approximated analytical
solutions in color dipole model, have been reported in last years
[21,22] with considerable phenomenological success. The analytical
methods of the unpolarized DGLAP evolution equations have been
discussed considerably in  Mellin and Laplace transformation
[23,24].\\
We present a modification of the DGLAP improved saturation model,
with respect to the Laplace transform technique by employing the
parametrization of proton structure function at leading-order up
to next-to-next-to-leading order (NNLO) approximations, which
preserves its behavior success in the low and high $Q^{2}$
regions. We show that GS  holds for the DGLAP improved model in a
wide kinematic region $rQ_{s}$. In next section, we introduce the
theoretical details of the model due to the Laplace transform
technique and discuss its qualitative features. We then derive the
dipole cross section with respect to the parametrization of
$F_{2}$ at LO up to NNLO approximations. In Section 3 we describe
our results and discuss their physical implications in comparison
with the GBW model. Section 4 contains conclusions.\\

\subsection{2. The Model}

An analytical expression for $F_{2}(x,Q^{2})$ has suggested by
authors in Ref. [25] which describes fairly well the available
experimental data on the reduced cross section in full accordance
with the Froissart predictions [26]. This parameterization
provides reliable structure function $F_{2}(x,Q^{2})$ according to
a combined fit of the H1 and ZEUS Collaborations data [27] in a
range of the kinematical variables $x$ and $Q^{2}$, $x{\leq}0.1$
and $0.15~\mathrm{GeV}^{2}<Q^{2}<3000~\mathrm{GeV}^{2}$, as
\begin{eqnarray}
F_{ 2}(x,Q^{2})& =& D(Q^{2})(1-
x)^{n}\sum_{m=0}^{2}A_{m}(Q^{2})L^{m},
\end{eqnarray}
and can be applied as well in analyses of ultra-high energy
processes with cosmic neutrinos. The effective parameters are
defined by the following forms
\begin{eqnarray}
D(Q^{2})=\frac{Q^{2}(Q^{2}+\lambda{M^{2}})}{(Q^{2}+M^{2})^{2}},~A_{0}(Q^{2})
=a_{00}+a_{01}L_{2},\nonumber\\
A_{i}(Q^{2})=\sum_{k=0}^{2}a_{ik}L_{2}^{k},~~i=(1,2),~~~~~~~~~~~
\end{eqnarray}
with the logarithmic terms $L$ as
\begin{eqnarray}
L=\ln\frac{1}{x}+L_{1},~~L_{1}=\ln(\frac{Q^{2}}{Q^{2}+\mu^{2}}),\nonumber\\
L_{2}=\ln(\frac{Q^{2}+\mu^{2}}{Q^{2}}),~~~~~~~~~~~~~
\end{eqnarray}
where the effective parameters $M$ and $\mu^{2}$ are the effective
mass and a scale factor, respectively. The additional parameters
with their statistical errors are given in Table I. According to
the DGLAP $Q^{2}$-evolution equation, the
 singlet and gluon distribution functions are related by the
 following form
\begin{eqnarray}
\frac{{\partial}F_{2}(x,Q^{2})}{{\partial}{\ln}Q^{2}}&=&-\frac{a_{s}(Q^{2})}{2}[P_{qq}(x){\otimes}F_{2}(x,Q^{2})\nonumber\\
&&+<e^{2}>P_{qg}(x){\otimes}xg(x,Q^{2})],\nonumber\\
\end{eqnarray}
where
\begin{eqnarray}
P_{a,b}(x)=P_{a,b}^{(0)}(x)
+a_{s}(Q^{2})\widetilde{P}_{a,b}^{(1)}(x)+a_{s}^{2}(Q^{2})\widetilde{P}_{a,b}^{(2)}(x)
\end{eqnarray}
and
\begin{eqnarray}
\widetilde{P}_{ab}^{(n)}(x)={P}_{ab}^{(n)}(x)+[C_{2,s}+C_{2,g}+...]\otimes
{P}_{ab}^{(0)}(x)+... .\nonumber
\end{eqnarray}
The quantities $\widetilde{P}_{ab}$$^{,}s$ are expressed via the
known splitting  and Wilson coefficient functions  in the
literatures [28,29] and $a_{s}(Q^{2})=\alpha_{s}(Q^{2})/4\pi$.\\
One can substantially simplify the calculations by considering
Eq.(13) in the space of  Laplace transform techniques, and taking
advantage of the fact the convolution form
$f_{1}(x){\otimes}f_{2}(x)$ in $x$ space becomes merely a product
of individual Laplace transforms of the corresponding functions in
the Laplace space. By considering the variable definitions
$\upsilon{\equiv}\ln(1/x)$ and $w{\equiv}\ln(1/z)$, Eq.(13) reads
as
\begin{eqnarray}
\frac{\partial{\mathcal{\widehat{F}}_{2}(\upsilon,Q^{2})}}{\partial{\ln}Q^{2}}&=&\int_{0}^{\upsilon}[\mathcal{\widehat{F}}_{2}(\upsilon,Q^{2})
\mathcal{\widehat{H}}^{(\varphi)}_{2,s}(a_{s}(Q^{2}),\upsilon-w)\\
&&+<e^{2}>\mathcal{\widehat{G}}(\upsilon,Q^{2})
\mathcal{\widehat{H}}^{(\varphi)}_{2,g}(a_{s}(Q^{2}),\upsilon-w)]dw,\nonumber
\end{eqnarray}
where
\begin{eqnarray}
\frac{\partial{\mathcal{\widehat{F}}_{2}(\upsilon,Q^{2})}}{\partial{\ln}Q^{2}}&{\equiv}&
\frac{{\partial}F_{2}(e^{-\upsilon},Q^{2})}{\partial{\ln}Q^{2}},\nonumber\\
\mathcal{\widehat{G}}(\upsilon,Q^{2})&{\equiv}&G(e^{-\upsilon},Q^{2}),\nonumber\\
\mathcal{\widehat{H}}^{(\varphi)}(a_{s}(Q^{2}),\upsilon)&{\equiv}&e^{-\upsilon}\widehat{P}_{a,b}^{(\varphi)}(a_{s}(Q^{2}),\upsilon).\nonumber
\end{eqnarray}
Here $\phi$ denotes the order in running coupling
$\alpha_{s}(Q^{2})$ and
$$ P_{a,b} ^{(\varphi)}
(a_{s},x)=\sum_{\phi=0}^{\varphi}a_{s}^{\phi+1}(Q^{2})P_{a,b}^{(\phi)}(x).
$$
The Laplace transform of
$\mathcal{\widehat{H}}(a_{s}(Q^{2}),\upsilon)$$^{,}s$
 are given by the following
 forms
\begin{eqnarray}
\Phi_{f}^{(\varphi)}(a_{s}(Q^{2}),s)&{\equiv}&
{\mathcal{L}}[\mathcal{\widehat{H}}^{(\varphi)}_{2,s}(a_{s}(Q^{2}),\upsilon);s]\nonumber\\
&&=\int_{0}^{\infty}\mathcal{\widehat{H}}^{(\varphi)}_{2,s}(a_{s}(Q^{2}),\upsilon)e^{-s\upsilon}d\upsilon,\nonumber\\
 \Theta_{f}^{(\varphi)}(a_{s}(Q^{2}),s)&{\equiv}&{\mathcal{L}}[\mathcal{\widehat{H}}^{(\varphi)}_{2,g}(a_{s}(Q^{2}),\upsilon);s]\nonumber\\
 &&=\int_{0}^{\infty}\mathcal{\widehat{H}}^{(\varphi)}_{2,g}(a_{s}(Q^{2}),\upsilon)e^{-s\upsilon}d\upsilon.\nonumber
\end{eqnarray}
We know that the Laplace transforms of the convolution factors are
simply the ordinary products of the Laplace transforms of the
factors. Therefore, Eq.(15) in the Laplace space $s$ reads as
\begin{eqnarray}
\frac{\partial{f_{2}(s,Q^{2})}}{\partial{\ln}Q^{2}}&=&
\Phi_{f}^{(\varphi)}(a_{s}(Q^{2}),s)f_{2}(s,Q^{2})\\
&&+<e^{2}>\Theta_{f}^{(\varphi)}(a_{s}(Q^{2}),s)g(s,Q^{2}),\nonumber
\end{eqnarray}
where
\begin{eqnarray}
{\mathcal{L}}[\mathcal{\widehat{F}}_{2}(\upsilon,Q^{2});s]&=&f_{2}(s,Q^{2}),\nonumber\\
{\mathcal{L}}[\mathcal{\widehat{G}}(\upsilon,Q^{2});s]&=&g(s,Q^{2}).\nonumber
\end{eqnarray}
The gluon distribution into the parametrization of the proton
structure function and its derivative with respect to ${\ln}Q^{2}$
in $s$-space in Eq.(16) is given by the following form
\begin{eqnarray}
g^{(\varphi)}(s,Q^{2})&=&k^{(\varphi)}(a_{s}(Q^{2}),s)Df_{2}(s,Q^{2})\nonumber\\
&&-h^{(\varphi)}(a_{s}(Q^{2}),s)f_{2}(s,Q^{2}),
\end{eqnarray}
where
\begin{eqnarray}
Df_{2}(s,Q^{2})&=&{\partial{f_{2}(s,Q^{2})}}/{\partial{\ln}Q^{2}},\nonumber\\
k^{(\varphi)}(a_{s}(Q^{2}),s)&=&1/(<e^{2}>\Theta^{(\varphi)}_{f}(a_{s}(Q^{2}),s)),\nonumber\\
h^{(\varphi)}(a_{s}(Q^{2}),s)&=&\Phi^{(\varphi)}_{f}(a_{s}(Q^{2}),s)k^{(\varphi)}(a_{s}(Q^{2}),s).\nonumber
\end{eqnarray}
The coefficient functions $\Phi_{f}$ and $\Theta_{f}$ in the
Laplace space $s$ are given by:\\
$\bullet$ at LO approximation\\
\begin{eqnarray}
\Theta_{f}^{(0)}(a_{s},s)&=&2n_{f}a_{s}(Q^{2}){\Big{[}}\frac{1}{1+s}-\frac{2}{2+s}+\frac{2}{3+s}{\Big{]}},\\
\Phi_{f}^{(0)}(a_{s},s)&=&a_{s}(Q^{2}){\Big{[}}4-\frac{8}{3}(\frac{1}{1+s}+\frac{1}{2+s}+2S_{1}(s){\Big{]}},\nonumber
\end{eqnarray}
Here $S_{1}(s)=\psi(s+1)+\gamma_{E}$, where $\psi(x)$ is the
digamma function and $\gamma_{E}=0.5772156
. . .$ is Euler constant.\\
The explicit expressions for the NLO and NNLO kernels in $s$ space
are rather cumbersome; therefore, we recall that we are interested
in investigation of the kernels in small $x$ [30,31], as\\
$\bullet$ at NLO approximation\\
\begin{eqnarray}
\Theta_{f}^{(1)}(a_{s},s)&{\simeq}&\Theta_{f}^{(0)}(s)+a^{2}_{s}(Q^{2})C_{A}T_{f}{\Big{[}}\frac{40}{9s}{\Big{]}},\\
\Phi_{f}^{(1)}(a_{s},s)&{\simeq}&\Phi_{f}^{(0)}(s)+a^{2}_{s}(Q^{2})C_{F}T_{f}{\Big{[}}\frac{40}{9s}{\Big{]}},\nonumber
\end{eqnarray}
and\\
$\bullet$ at NNLO approximation\\
\begin{eqnarray}
\Theta_{f}^{(2)}(a_{s},s)&{\simeq}&\Theta_{f}^{(1)}(s)+a^{3}_{s}(Q^{2}){\Big{\{}}n_{f}{\Big{[}}-\frac{1268.300}{s}+\frac{896}{3s^{2}}{\Big{]}}\nonumber\\
&&+n^{2}_{f}{\Big{[}}\frac{1112}{243s}{\Big{]}}{\Big{\}}},\\
\Phi_{f}^{(2)}(a_{s},s)&{\simeq}&\Phi_{f}^{(1)}(s)+a^{3}_{s}(Q^{2}){\Big{\{}}n_{f}{\Big{[}}-\frac{506}{s}+\frac{3584}{27s^{2}}{\Big{]}}\nonumber\\
&&+n^{2}_{f}{\Big{[}}\frac{256}{81s}{\Big{]}}{\Big{\}}}.\nonumber
\end{eqnarray}
The standard representation for QCD couplings in LO up to NNLO
(within the $\mathrm{\overline{MS}}$-scheme) approximations are
defined by
\begin{eqnarray}
\alpha_{s}^{\mathrm{LO}}(t)&=&\frac{4\pi}{\beta_{0}t},\nonumber\\
\alpha^{\mathrm{NLO}}_{s}(t)&=&\frac{4\pi}{\beta_{0}t}\Big{[}1
-\frac{\beta_{1}}{\beta_{0}^{2}}\frac{\ln{t}}{t}\Big{]},\nonumber\\
\alpha^{\mathrm{NNLO}}_{s}(t)&=&\frac{4\pi}{\beta_{0}t}\Big{[}1
-\frac{\beta_{1}}{\beta_{0}^{2}}\frac{\ln{t}}{t}\nonumber\\
&&+\frac{1}{\beta_{0}^{3}t^{2}}\bigg{\{}
\frac{\beta_{1}^{2}}{\beta_{0}}(\ln^{2}t-\ln{t}-1)+\beta_{2}\bigg{\}}
\Big{]},\nonumber
\end{eqnarray}
where $\beta_{0}$, $\beta_{1}$ and $\beta_{2}$ are the one, two
and three loop correction to the QCD $\beta$-function and
$t=\ln\frac{Q^{2}}{\Lambda^{2}}$, $\Lambda$ is the QCD cut-off
parameter.\\
Now the inverse Laplace transforms of Eq.(17) can be easily
performed by the following form as
\begin{eqnarray}
\widehat{G}^{(\varphi)}(\upsilon,Q^{2})&{\equiv}&{\mathcal{L}}^{-1}[g^{(\varphi)}(s,Q^{2})(s,Q^{2});\upsilon]\nonumber\\
&&={\mathcal{L}}^{-1}[k^{(\varphi)}(a_{s}(Q^{2}),s)Df_{2}(s,Q^{2})\nonumber\\
&&-h^{(\varphi)}(a_{s}(Q^{2}),s)f_{2}(s,Q^{2});\upsilon],
\end{eqnarray}
where the inverse transform of a product to the convolution of the
original functions, giving [32]
\begin{eqnarray}
{\mathcal{L}}^{-1}[f(s){\times}h(s);\upsilon]&=&\int_{0}^{\upsilon}\widehat{F}
(w)\widehat{H}(\upsilon-w)dw.\nonumber
\end{eqnarray}
The result for the color dipole cross section at scale $\mu^{2}$
is
\begin{eqnarray}
\sigma^{(\varphi)}_{\mathrm{dip}}=\sigma_{0}\{1-\exp(-\frac{\pi^{2}r^{2}\alpha_{s}(\mu^{2})G^{(\varphi)}(\widetilde{x},\mu^{2})}{3\sigma_{0}})\},
\end{eqnarray}
where
\begin{eqnarray}
{G}^{(\varphi)}(x,\mu^{2})&=&\int_{x}^{1}\Big{[}{DF}_{2}
(\ln\frac{1}{y},\mu^{2}){k}^{(\varphi)}(a_{s}(\mu^{2}),\ln\frac{y}{x})\nonumber\\
&&-{F}_{2}
(\ln\frac{1}{y},\mu^{2}){h}^{(\varphi)}(a_{s}(\mu^{2}),\ln\frac{y}{x})\Big{]}\frac{dy}{y}.\nonumber\\
\end{eqnarray}
We therefore obtained an explicit solution for the color dipole
cross section $\sigma_{\mathrm{dip}}(x,r)$ in terms of the
parametrization of $F_{2}(x,\mu^{2})$ and its derivative with
respect to ${\ln}\mu^{2}$ at LO up to NNLO approximations due to
the form of kernels.\\

\subsection{3. Numerical Results}

The effective parameters in the GBW model have been extracted from
a fit of the HERA data  according to Ref. [1]:
\begin{eqnarray}
\sigma_{0}=23~ \mathrm{mb},~ \lambda=0.288,~
x_{0}/10^{-4}=3.04,\nonumber\\
C=0.38,
~\mu_{0}^{2}=1.73~\mathrm{GeV}^{2}~~~~~~~~~~~~~~~~~~~\nonumber
\end{eqnarray}
We have calculated the $r$-dependence, at low $x$, of the ratio
$\sigma_{\mathrm{dip}}/\sigma_{0}$ (i.e., Eq.(22)) in the LO up to
NNLO, approximations. Results of calculations and comparison with
the GBW model [1] are presented in Figs.1-3, where the circle-dot
lines correspond to the extracted
$\sigma_{\mathrm{dip}}/\sigma_{0}$ in the LO up to NNLO
approximations, respectively.
\begin{figure}
\includegraphics[width=0.5\textwidth]{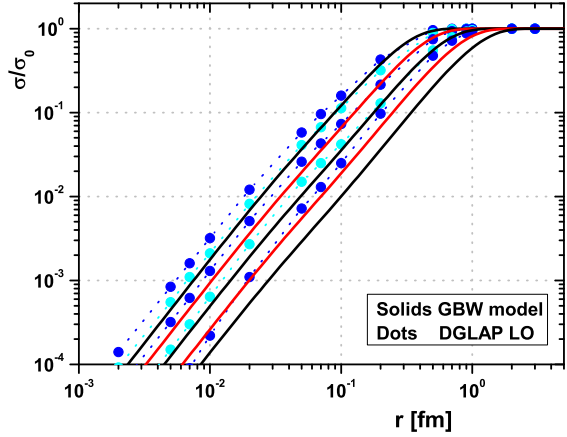}
\caption{The extracted ratio $\sigma_{\mathrm{dip}}/\sigma_{0}$ as
a function $r$ for $x=10^{-6}..10^{-2}$ (curves from left to
right, respectively) from the parameterization of $F_{2}$ within
the LO approximation (circle-dot curves), Eq.(22), compared with
the GBW model (solid curves), Eq.(3).}\label{Fig1}
\end{figure}
\begin{figure}
\includegraphics[width=0.5\textwidth]{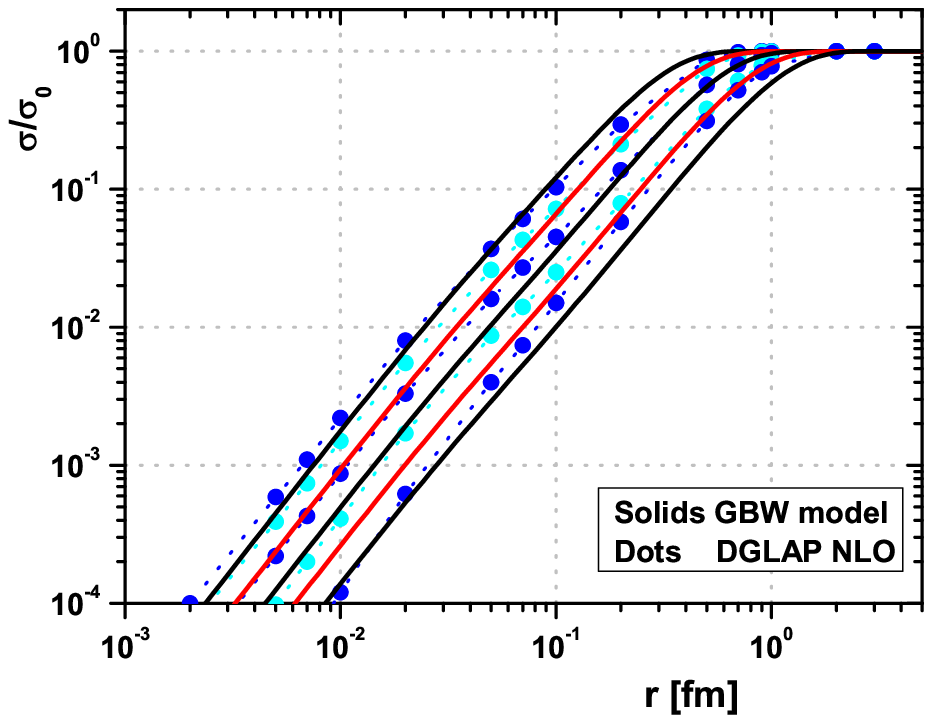}
\caption{The same as Fig.1 within the NLO approximation.
}\label{Fig2}
\end{figure}
\begin{figure}
\includegraphics[width=0.5\textwidth]{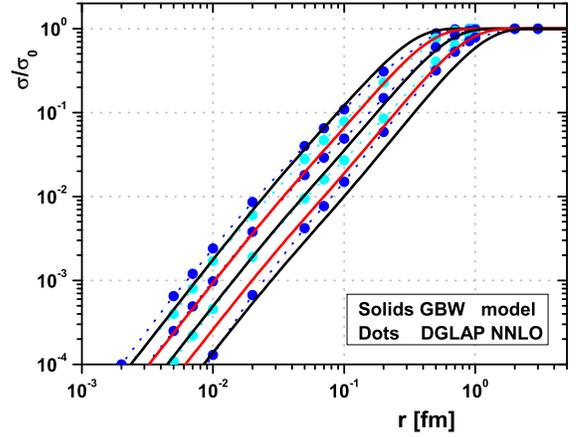}
\caption{The same as Fig.1 within the NNLO
approximation.}\label{Fig3}
\end{figure}
Calculations have been performed at  the Bjorken variable $x$ to
vary in the interval $x=10^{-6}..10^{-2}$. The DGLAP improved
model due to the parameterization of $F_{2}(x,Q^{2})$ giving a
good description of the ratio $\sigma_{\mathrm{dip}}/\sigma_{0}$
in comparison with the GBW saturation model at low $x$ in a wide
range of the momentum transfer $Q^{2}$. Figures 1-3 clearly
demonstrates that the extraction procedure provides correct
behaviors of the extracted $\sigma_{\mathrm{dip}}/\sigma_{0}$
within the LO up to NNLO approximations. At low and high $Q^{2}$
the extracted values of $\sigma_{\mathrm{dip}}/\sigma_{0}$ are in
a good agreement with the GBW saturation model. We observe that
the NNLO corrections  are  in a very good agreement with the GBW
model in comparison with the LO and NLO corrections in a wide
range of $r$. We see that the two results (the GBW and DGLAP
improved models) overlap in small and large values of $r$, where
the gluon distribution obtained from the parametrization of the
proton structure function plays a significant role
in the evolution of the gluon distribution.\\
\begin{figure}
\includegraphics[width=0.5\textwidth]{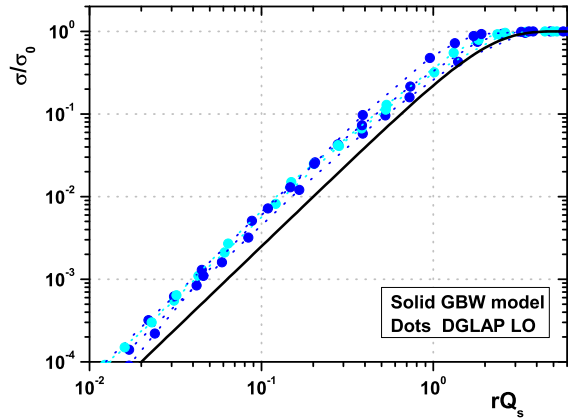}
\caption{The extracted ratio
$\sigma_{\mathrm{dip}}(rQ_{s}(x))/\sigma_{0}$ as a function
$rQ_{s}$ for $x=10^{-6}..10^{-2}$ from the parameterization of
$F_{2}$ within the LO approximation (circle-dot curves) merges
into one line due to the geometric scaling and compared with the
GBW model (solid curve).}\label{Fig4}
\end{figure}
\begin{figure}
\includegraphics[width=0.5\textwidth]{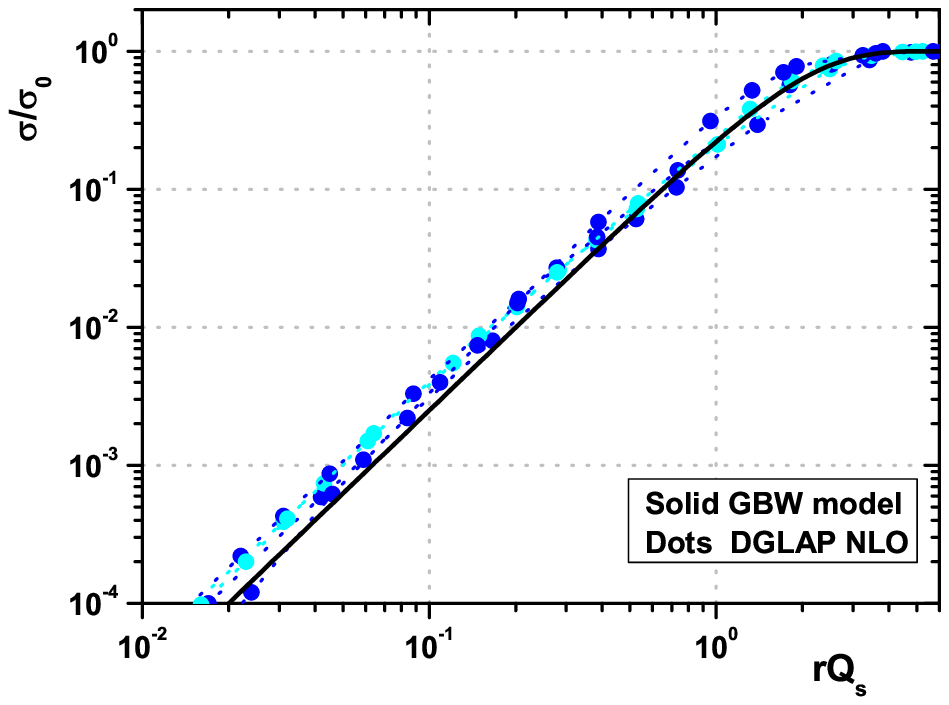}
\caption{The same as Fig.4 within the NLO
approximation.}\label{Fig5}
\end{figure}
\begin{figure}
\includegraphics[width=0.5\textwidth]{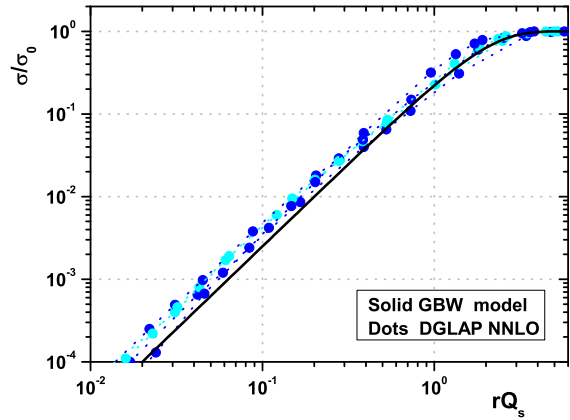}
\caption{The same as Fig.4 within the NNLO
approximation.}\label{Fig6}
\end{figure}
A particular interests present the ratio
$\sigma_{\mathrm{dip}}/\sigma_{0}$ defined by the scaling variable
$rQ_{s}$ where all the curves in the GBW model merge into one
solid line. In Figs.4-6 we have shown that the ratio
$\sigma_{\mathrm{dip}}(x,r)/\sigma_{0}$ has a property of
geometric scaling as
$\sigma_{\mathrm{dip}}(x,r)=\sigma_{\mathrm{dip}}(rQ_{s}(x))$. The
results of the DGLAP improved saturation model due to the
parametrization of the proton structure function have become a
function of a single variable, $rQ_{s}$, for all values of $r$ and
$x$ at LO up to NNLO approximations in figures 1-3 respectively.
From Figure 6 one can infer that the NNLO results essentially
improve the agreement with the geometric scaling in the GBW model
in comparison with the LO and NLO calculations. The geometric
scaling in the dipole cross sections in these calculations is
visible in a wide range of $rQ_{s}$ at LO up to NNLO
approximations. In these figures we observe that the violation
between the geometric scaling of our results and GBW model for low
$rQ_{s}$ is clearly visible. In this region, the violations are
rather small and can be covered by the statistical errors in the
parametrization of the proton structure function and its
derivative. In Fig.7, the $rQ_{s}$ dependence of the ratio
$\sigma_{\mathrm{dip}}(rQ_{s}(x))$ at $x=10^{-4}$ compared with
the GBW saturation model. The error bands illustrated in this
figure are the statistical errors in the parametrization of
$F_{2}$ and its derivative, where the fit parameter errors are
shown in Table I. As can be seen from the related figures, the
ratio results with respect to the Laplace transform method are
consistent with the geometric scaling at low and large values of
$rQ_{s}$.
\begin{figure}
\includegraphics[width=0.5\textwidth]{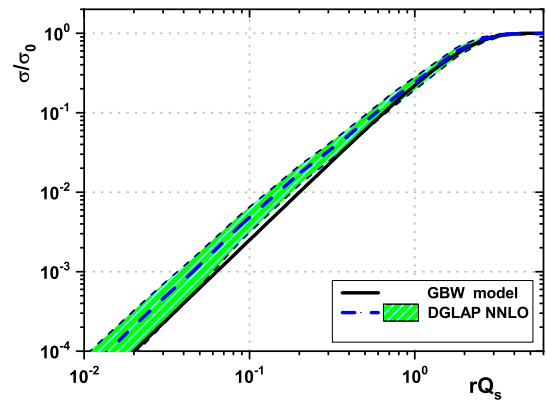}
\caption{The extracted ratio
$\sigma_{\mathrm{dip}}(rQ_{s}(x))/\sigma_{0}$ as a function
$rQ_{s}$ for $x=10^{-4}$ within the NNLO approximation (dashed
curves), accompanied with the statistical errors in the
parametrization of $F_{2}$ and its derivative, compared with the
GBW model (solid curve).}\label{Fig7}
\end{figure}
Summarising, the essential elements of the GBW model, the
saturation scale and geometric scaling, are preserved  in the
DGLAP improved dipole cross section when the gluon distribution
function is derivative from the parametrization of the proton
structure function and its derivative due to the Laplace
transforms method
in a wide range of the variables $r$ and $rQ_{s}$ respectively.\\

\subsection{4. Conclusions}

In conclusion, we have presented a certain theoretical model at LO
up to NNLO approximations to describe the color dipole cross
section based on the Laplace transforms method at small values of
$x$. Indeed, there are various methods to consider the color
dipole model to obtain $\sigma_{\mathrm{dip}}$, and in this paper,
we have shown that the method of the Laplace transform technique
is also the reliable and alternative scheme to obtain the color
dipole cross section, analytically. A detailed analysis has been
performed to find an analytical solution of the color dipole cross
section into the parametrization of $F_{2}(x,Q^{2})$ and its
derivative of the proton structure function with respect to
${\ln}Q^{2}$ at LO up to NNLO approximations. We used the DGLAP
improved model of the dipole cross section with saturation in
which the parameterization of the proton structure function is
used. The results according to the saturation scale and geometric
scaling are consistent with the GBW saturation model in a wide
range of $r$ and $rQ_{s}$, respectively. With considering the
statistical errors due to the effective parameters, the NNLO
results  give  a reasonable data description in comparison with
the other models. Indeed, the small dipole size part of the dipole
cross section is improved in comparison with the DGLAP improved
model which is based on  the evolution of gluon density in this
region. As a summary, we have analyzed the dipole cross section at
low values of $x$ and shown that the geometric scaling holds  for
the DGLAP improved model if the gluon distribution is defined by
the paramterization of the proton structure function and this is
comparable with the GBW model curve in
the whole region $rQ_{s}$.\\

\subsection{ACKNOWLEDGMENTS}
The author is grateful to Razi University for the financial
 support of this project and would like to thank A.M.Stasto for carefully reading the paper and for critical notes.\\


\begin{table}[h]
\caption{ The effective Parameters at low $x$ for
$0.15~\mathrm{GeV}^{2}<Q^{2}<3000~\mathrm{GeV}^{2}$ provided by
the following values. The fixed  parameters are defined by the
Block-Halzen fit to the real photon-proton cross section as
$M^{2}=0.753 \pm 0.068~ \mathrm{GeV}^{2}$ and $\mu^2 = 2.82 \pm
0.290~ \mathrm{GeV}^{2}$.}
\begin{tabular} {cccc}
\toprule \\  \multicolumn{2}{c}{parameters \quad \quad \quad ~~~~~~~~~~~~~~~~value}    \\ &&&\\ \hline \\ &&&\\
$a_{00}$& \quad  $2.550\times 10^{-1}~\pm 1.60\times10^{-2}$ \\

$a_{01}$& \quad  $1.475\times 10^{-1}~\pm 3.025\times10^{-2}$\\&&&\\

  $a_{10} $  &   \quad  $8.205\times 10^{-4}~~  \pm  4.62\times10^{-4} $  \\

  $a_{11} $  &   \quad   $-5.148\times 10^{-2}\pm 8.19\times10^{-3}$  \\

  $a_{12}$   &    \quad  $-4.725\times 10^{-3}\pm 1.01\times10^{-3}$   \\  &&&\\

 $a_{20}$   &   \quad   $2.217\times 10^{-3}\pm 1.42\times10^{-4} $ \\

 $a_{21}$   &   \quad   $1.244\times 10^{-2}\pm 8.56\times10^{-4}$  \\

 $a_{22}$    &    \quad  $5.958\times 10^{-4}\pm 2.32\times10^{-4} $ \\ &&& \\

$n$& \quad  $11.49\pm 0.99$ & &\\

$\lambda$& \quad  $2.430~\pm 0.153$ & &\\

$\chi^{2}(\mathrm{goodness~ of~ fit})$ &  \quad  $0.95$ & &\\

\hline

\end{tabular}
\end{table}

\section{References}
1. K. Golec-Biernat and S.Sapeta, JHEP {\bf03},
102 (2018).\\
2. I. Abt et al., Phys. Rev. D {\bf96},  014001 (2017).\\
3. F. D. Aaron et al., [H1 and ZEUS Collaborations],  JHEP
{\bf01}, 109 (2010); H. Abramowicz et al., [H1 and ZEUS
Collaborations], Eur. Phys. J. C{\bf75}, 580 (2015).\\
4. Yu.L.Dokshitzer, Sov.Phys.JETP {\textbf{46}}, 641(1977);
G.Altarelli and G.Parisi, Nucl.Phys.B \textbf{126}, 298(1977);
V.N.Gribov and L.N.Lipatov, Sov.J.Nucl.Phys. \textbf{15},
438(1972).\\
5. V.S.Fadin, E.A.Kuraev and L.N.Lipatov, Phys.Lett.B \textbf{60},
50(1975); L.N.Lipatov, Sov.J.Nucl.Phys. \textbf{23}, 338(1976);
I.I.Balitsky and L.N.Lipatov, Sov.J.Nucl.Phys. \textbf{28}, 822(1978).\\
6. M.Kuroda and D.Schildknecht, Phys.Lett. B{\bf670}, 129(2008); Phys.Rev. D{\bf96}, 094013(2017); Int. J. Mod. Phys. A{\bf31}, 1650157 (2016).\\
7. Amir H.Rezaeian and I.Schmidt, Phys.Rev. D{\bf88}, 074016 (2013).\\
8. K.Golec-Biernat and  M.W$\ddot{\mathrm{u}}$sthoff, Phys. Rev. D
{\bf59},  014017 (1999); Phys. Rev. D {\bf60},  114023 (1999).\\
9. E.Iancu,K.Itakura and S.Munier, Phys.Lett.B {\bf590}, 199
(2004).\\
10. J.Bartels, K.Golec-Biernat and H.Kowalski, Phys. Rev.
D{\bf66}, 014001 (2002); Acta Phys.Polon.B {\bf33}, 2853 (2002).\\
11. J.R.Forshaw and G.Shaw, JHEP {\bf12},
052 (2004).\\
12. K.Golec-Biernat, J.Phys.G {\bf28}, 1057 (2002); Acta Phys.Polon.B {\bf33}, 2771 (2002).\\
13. H.Kowalski and D.Teaney, Phys. Rev.
D{\bf68}, 114005 (2003).\\
14. G. Soyez, arXiv: 0705.3672 (2007); M.V.T. Machado, arXiv:
0512264 (2006); J. T. de Santana Amaral et al., arXiv:0612091
(2006).\\
15. B. Ducloue et al., arXiv:1912.09196 (2019).\\
16. K.Golec-Biernat and S. Sapeta, Phys. Rev. D{\bf74}, 054032
(2006).\\
17. I. Balitsky, Nucl. Phys. B{\bf463}, 99 (1996); Y. V.
Kovchegov,
Phys. Rev. D{\bf60},  034008(1999); Phys. Rev. D{\bf61}, 074018 (2000).\\
18. A.M.Stasto, K.Golec-Biernat and J.Kwiecinski,
Phys.Rev.Lett.{\bf86}, 596 (2001).\\
19. E. Iancu and R. Venugopalan, arXiv:hep-ph/0303204.\\
20. Y.S.Jeong, C.S.Kim, M.V.Luu and M.H.Reno, JHEP {\bf11}, 025
(2014).\\
21. Z.Jalilian and G.R.Boroun, Phys.Lett.B {\bf773}, 455 (2017);
Z.Jalilian and G.R.Boroun, Chin.Phys.C {\bf45}, 023101 (2020);
B.Rezaei and G.R.Boroun, Phys.Rev.C {\bf101}, 045202 (2020).\\
22.G.R.Boroun and B.Rezaei, Nucl.Phys.A {\bf990}, 244 (2019);
G.R.Boroun, Eur.Phys.J.A {\bf57}, 219 (2021); G.R.Boroun and
B.Rezaei, Phys.Rev.C {\bf103}, 065202 (2021).\\
23. L.P.Kaptari, A.V.Kotikov, N.Yu.Chernikova, and P.Zhang, Phys.Rev.D {\bf99}, 096019 (2019).\\
24. G.R.Boroun and B.Rezaei, Phys.Rev.D {\bf105}, 034002 (2022).\\
25. M. M. Block, L. Durand and P. Ha, Phys. Rev.{\bf D 89}, no. 9,
094027 (2014).\\
26. M. Froissart, Phys. Rev. {\bf123}, 1053 (1961).\\
27. F.D. Aaron et al., [H1 and ZEUS Collaborations], JHEP
{\bf1001}, 109 (2010).\\
28. J. Blumlein, V. Ravindran and W. van Neerven, Nucl. Phys. B
\textbf{586}, 349(2000); S.Catani and F.Hautmann,
Nucl.Phys.B{\bf427}, 475(1994).\\
29. D.I.Kazakov and A.V.Kotikov, Phys.Lett.B{\bf291}, 171(1992);
E.B.Zijlstra and W.L.van Neerven, Nucl.Phys.B{383}, 525(1992).\\
30. A. Vogt, S. Moch and J.A.M. Vermaseren, Nucl.Phys.B {\bf691},
129 (2004).\\
31. W.L. van Neerven and A.Vogt, Phys.Lett.B {\bf490}, 111
(2000).\\
32. M.M. Block, L. Durand, and D.W. McKay, Phys. Rev. D {\bf79},
014031 (2009).\\


\end{document}